\documentstyle[12pt]{article}
\newcommand{\beq}{\begin{equation}}
\newcommand{\eeq}{\end{equation}}
\begin{document}    \begin{flushright}  PITHA 96/16 \\
 gr-qc/9605038 \end{flushright}  \begin{center} {\LARGE On the Quantum
Levels  \\[0.1cm] of Isolated Spherically Symmetric \\[0.2cm]
 Gravitational Systems} \\[2cm]
{\Large H.A. Kastrup\footnote{E-Mail: kastrup@physik.rwth-aachen.de} \\
Institute for Theoretical Physics\\ RWTH Aachen \\[0.2cm] 52056 Aachen, Germany}
\end{center} \vspace*{1.5cm}
  {\large \bf Abstract} \\[0.2cm]   The known canonical quantum theory of a
  spherically symmetric pure (Schwarzschild) gravitational system describes
  isolated black holes by plane waves $\exp(-iMc^2\tau/\hbar)$
  with respect to their continuous masses $M$
   and  the proper time $\tau$ of observers at spatial infinity. \\On the
    other hand
   Bekenstein and Mukhanov postulated discrete mass levels for such black
   holes in the spirit of the Bohr-Sommerfeld quantisation in atomic physics.
   \\ The two approaches can be related by postulating periodic boundary
   conditions in time for the plane waves and by identifying the period
   $\Delta$ in real time with the period $\Delta_H= 8\pi G\,M/c^3$ in Euclidean
   time. This yields the mass spectrum $M_n=(1/2)\sqrt{n}\,m_P\,,
   n=1,2,\cdots~$. \newpage
   \section{Introduction} The
  quantum theory of black holes coupled to matter has been studied intensively
  during the last years. Most of the time one starts with a quantised matter
  field coupled to the classical geometry of a black hole and then asks for
  the backreaction of this coupling to the geometry (for two recent
  introductions see e.g.\ the reviews \cite{wa1}
   and \cite{br}). The geometry
  itself is generally not quantised, because an analysis of the totally
   quantised
  coupled systems  meets so severe difficulties that such an approach up to
  now has resisted a breakthrough, at least in 4 spacetime dimensions (for a
  recent review of the  state of the art see ref. \cite{is}). \\ There is,
   however,
  another possible approach to the problem, which starts from the quantum theory
  of an  isolated spherically symmetric gravitational system and then studies
  the coupling of
  such a quantum system to (quantised) matter, at least approximately. This
  approach might have the advantage that a considerable part of
  the vast gravitational gauge
  structure has already been dealt with.  It is still in its
  infancy and only the first step, namely the canonical quantisation of
   spherically
  symmetric pure gravity has been achieved \cite{kt} (Thiemann and myself
   used Ashtekar's
  framework, Kucha\v{r} \cite{ku} the usual geometrodynamical one). \\
  One the other hand, long before
  this systematically derived quantum theory of  isolated Schwarzschild
  black holes the quantisation of its energy (mass) levels had been postulated
  heuristically first by Bekenstein and later by Mukhanov and others in the
  spirit of the Bohr-Sommerfeld quantisation rules in atomic physics. The idea
  and the results of this approach have been discussed again in a recent
   letter by
  Bekenstein and Mukhanov \cite{bm} leading to a lively response by Ashtekar
  and his school \cite{as} for which "area"  is one of the basic observables
   in their
  quantisation program in terms of loop variables. \\ In the following I want
  to indicate how the main features of Bekenstein's and Mukhanov's approach
  may be related to our (and Kucha\v{r}'s) "canonical" results. The arguments
   are
  quite straightforward and perhaps too simple, but probably worth  discussing!
  \section{Summary of the canonical and the Bekenstein-Mukhanov quantisation
  of black holes}In the following I recall only the most essential results
  needed for discussing the relation between the two quantum theoretical
  treatments of black
  holes mentioned above. All the details can be found in refs. [4], [5] and [6]
   and the
  literature quoted there. \\  The rotational symmetric line element is \beq
  ds^2 = -(N(r,t)dt)^2 + q_r(r,t)(dr+N^r(r,t)dt)^2 + R^2(r,t)d\Omega^2~. \eeq
  The velocity of light $c$, here put equal to 1, will be restored explicitely
   below.
  The Schwarzschild variables $r,t$ are meant to represent their
   extensions to the Kruskal
  manifold, too. So $r$ may run from $-\infty$ to $+\infty$ etc.. $N$ is the
  lapse and $N^r$ the radial shift function, $d\Omega^2=d\theta^2 +
  \sin^2\theta d\phi^2$. The manifolds are assumed to be asymptotically flat.
  \\ It is essential  not to start with the Schwarzschild gauge
  $R(r,t)=r, N^r=0$,  because otherwise  one misses
  essential elements of the canonical structure [4,5]. Solving the classical
   constraints
  leads to just one canonical pair of observables - in the sense of Dirac -,
  namely the mass $M(\ldots)$ (the dots indicate canonical variables in the
  original phase space) and a
  canonically conjugate time functional $T[\ldots]$ which may be represented
   as  \begin{eqnarray} T[q_r,R]&=& 2\int_{\Sigma}(1-2MG/R)^{-1}w(q_r,R)~,\\
   & & w(q_r,R)=[(R')^2-q_r(1-2MG/R)]^{1/2}~, \nonumber\end{eqnarray}  where
   $R'=dR/dr$ and where $\Sigma$ means a 1-dimensional "radial"
   (Cauchy) surface which extends from $r\rightarrow -\infty$ to
   $r\rightarrow +\infty$. $T$ and $M$ obey the Poisson brackets \beq
   \{T,M\}=1 \eeq among themselves and the following ones with the total
   (unreduced!) Hamiltonian $H_{tot}$ \begin{eqnarray} \dot{T}&=&\{T,H_{tot}\}=
   N_+(t)+N_-(t)~,\\ \dot{M}&=&\{M,H_{tot}\}=0~, \end{eqnarray}where
   $\dot{X}\equiv dX/dt$. Here $N_-(t)=
   N(r\rightarrow -\infty,t), N_+(t)=N(r\rightarrow +\infty,t)$ (I assume
   that there are 2 spatial asymptotic "ends" of the manifold like for the
   Kruskal one. Compared to ref.\ [4] I have divided the Hamiltonian by 2).
   The first of the last two equations shows the meaning of
   $T$: If $\tau_+, \tau_-$ are the proper times at the two spatial
   infinities,
   then $\dot{\tau}_+=N_+(t), \dot{\tau}_-=-N_-(t)$. Thus the functional $T$
  represents an observable "time" and the quantity
     \beq \delta =T-(
   \tau_+-\tau_-)\eeq has a vanishing Poisson bracket with $H_{tot}$ and
   therefore it is a constant! \\The functional $T$ is nonvanishing only if the
   shift $N^r$ does not vanish [4]. So it represents a measure for the slicing
   of
   spacetime into space and time relative to the static (Schwarzschild)
   slicing for which $T=0$! If one integrates in eq.\ (2) not till $r=\infty$
   but only till $r$ at the upper limit, then the resulting $T(t,r)$ can be
   interpreted as the Killing time associated with the system [5]. \\ If we
   restrict $H_{tot}$ to the surface of the phase space where the constraints
   vanish we get the reduced Hamiltonian \beq H_{red}= M(N_+ + N_-)~, \eeq which
   results from the nonvanishing (ADM) surface terms. Because of the Poisson
   brackets (3) we have $\{T,H_{red}\}=N_++N_-$ in agreement with the relation
   (4). \\ All this shows that  in the case of spherically symmetric pure
   gravity the complete elimination of  the gauge degrees of freedom leads to
    a 1+1
   dimensional integrable mechanical system with  the canonical coordinate
   $T$ and the conjugate momentum $M$ both of which form the reduced phase
   space! \\ Quantisation of the system conveniently starts from this reduced
   phase space (one could also adopt Dirac's approach, because the functional
   $T[\ldots]$ is annihilated by the constraints if one choses an appropriate
   operator ordering [4]): \\
    First we "promote" the quantities $E=Mc^2$ and $T$
   to operators $\hat{E}$ and $\hat{T}$ and the Poisson bracket (3) to the
   commutator \beq [\hat{E}, \hat{T}]=\hbar/i ~. \eeq Here we encounter a
   (physical) problem [4]: If $\hat{E}$ and $\hat{T}$ both are represented by
   selfadjoint operators then the spectrum $\{E\}$ of $\hat{E}$ has to be
    the whole
   real axis because the unitary operator $\exp(i\hat{T}\mu/\hbar)$, $\mu$ real,
    generates translations $E\rightarrow E+\mu$. This would lead to
    negative masses, naked
    singularities and unwanted instabilities. If, therefore, one wants the
    spectrum of $\hat{E}$ to be bounded from below, $E\ge 0$, then $\hat{T}$
    cannot be selfadjoint. A possible way out is to define \beq
    \hat{S}=\frac{1}{2}(\hat{E}\hat{T}+\hat{T}\hat{E})~,~~\mbox{with }
    [\hat{S},\hat{E}]=i\hbar~,\eeq because $\exp(i\hat{S}\beta/\hbar)$ just
    rescales $E$, $E\rightarrow  e^{\beta}E$.
    \\ Important is the Schroedinger equation: Consider an
    observer at $r\rightarrow +\infty$ with the reduced Hamilton operator \beq
    \hat{H}_{red}= N_+\hat{E} \eeq and with the proper time $\tau\equiv \tau_+,
    \dot{\tau}=N_+$. In the $M$-representation the wave function
    $\varphi(M,t)$ obeys the simple Schroedinger equation \begin{eqnarray}
     i\hbar
    \partial_t\varphi&=& \hat{H}_{red}\,\varphi =N_+\,Mc^2\varphi~,\mbox{  or}
    \nonumber \\ i\hbar\partial_{\tau}\varphi(M,\tau)&=& Mc^2\;
    \varphi(M,\tau)~,\end{eqnarray} with the solutions \beq
    \varphi(M,\tau)=\chi(M)\,e^{\displaystyle-\frac{i}{\hbar}Mc^2\,
    \tau}\eeq and the scalar
    product \beq (\varphi_1, \varphi_2) = \int_0^{\infty}dM
    \varphi_1^{\ast}\varphi_2~~. \eeq  Obviously the mass spectrum is
    continuous and if nothing happens the isolated gravitational sytem (black
    hole) just "sits" there,
    described by a plane wave in time!\\ Let me now turn to the quantisation
    scheme for black holes as discussed by Bekenstein and Mukhanov [6]: Here
    the area $A$ enclosed by the horizon of the black hole is the starting
    point, \beq A= 4\pi R_S^2~,~~ R_S= 2M\,G/c^2~. \eeq According to the
    old  Bohr-Sommerfeld
    quantisation rules in atomic physics finite 2-dimensional regions $B$
     with boundary
    $\partial B$ in phase should be quantised by the prescription \beq
    \frac{1}{2\pi}\oint_{\partial B}p\,dq =\frac{1}{2\pi} \int_{B}dp\wedge dq =
    n\, \hbar~,~ n=1,2,\ldots ~~. \eeq  This and related arguments lead to
    the assumption  that the area A (and therefore the mass $M$) are quantised
    accordingly: \beq A_n=16\,\pi\,G^2\ M_n^2/c^4 = const.\ n\,\hbar=
    \alpha\,n\, l_{P}^2~,~n=1,2,\ldots~~, \eeq where
    $l_{P}=(\hbar\,G/c^3)^{1/2}=1.6\cdot10^{-35}\mbox{ m}$ is
    Planck's length and $\alpha$ a dimensionless constant to be determined. We
    thus get the following quantisation of the mass: \beq M_n
    =\frac{1}{4}\sqrt{\frac{\alpha}{\pi}} \sqrt{n}\;m_{P}~,~
    m_{P}=\sqrt{c\hbar/G}=2.2\cdot 10^{-8}\mbox{ kg}= 1.2\cdot 10^{19} \mbox{
    GeV}/c^2~~. \eeq
    In order to determine the constant $\alpha$ the relation \beq
     S=\frac{k_Bc^3}{4 G\,\hbar}A +k_B a=
    \frac{k_B}{4}\frac{A}{l_{P}^2} + k_B a  \eeq between the entropy $S$ of a
    black hole and its area enclosed by the horizon can be used ($a$ is a
     normalisation
    constant for the entropy and $k_B$ is Boltzmann's constant). The constants
    $\alpha$ and $a$ can be specified as follows: Suppose g(n) is the
    degeneracy of the nth level. Then we can associate the (statistical)
    entropy $S=k_B\,\ln g(n)$ with it, or \beq g(n)=
    e^{\displaystyle S/k_B}=e^{\displaystyle \alpha\,n/4+a}~~. \eeq  Requiring
     $g(n=1)=1$ yields
    $a=-\alpha/4$ and since $g(n)$ has to be an integer one concludes
    $\alpha=4\ln k, k=2,3,\ldots$. If one now identifies $g(n)$ with the
    number of possible ways one can build up the nth level when one starts from
    "nothing" (n=0), then k=2, and therefore \beq g(n)=2^{n-1}~,~~ \alpha=
    4\ln2~,\eeq  so that finally \begin{eqnarray} A_n^{BM} &=&
     4\ln2~n\,l_{P}^2
    ~,~n=1,2, \ldots \\ S_n^{BM} &=&k_B\ln2~ (n-1)~,\\ M_n^{BM}
    &=&\frac{1}{2}\sqrt{\displaystyle
    \frac{\ln2}{\pi}}~\sqrt{n}\, m_{P}~~. \end{eqnarray} As the Hawking
    temperature at spatial infinity is given by \beq
    \beta_H=\frac{1}{k_BT_H}=\frac{8\pi\,G\,M}{\hbar\,c^3}~,\eeq the
    application of this formula to the quantised mass levels gives the rather
     strange result
    that the temperature is quantised, too! \section{Relating the canonical
    wave functions to quantised levels of the black hole} If one wants to
    relate the continuous mass values $M$ in the wave function (12) to the
    discrete mass levels (23) one has to impose additional conditions on
    $\varphi(M,\tau)$. The following two assumptions will achieve this:
    \\ 1. Suppose the plane wave (12) represents the gravitational system
     only for a
    finite time interval $\Delta \tau\equiv \Delta >0 $, because a collaps has
    terminated this state more or less abruptly. Formally this property can be
    implemented by the requirement that the wave function $\varphi(M,\tau)$
    obeys periodic boundary conditions, $\varphi(M,\Delta)=\varphi(M,0)$,
    which leads to \beq c^2 M\,\Delta=2\pi\hbar\, n,~~n=1,2,\ldots~~,\eeq
    where we have assumed that the mass is positive. The last relation makes
    the mass spectrum discontinuous in the same way as a spatial box with the
    corresponding boundary conditions makes the momentum for free particles
discrete. \\
2. The next problem is to determine the time interval $\Delta$. Intuitively
one expects that it should be larger than the time the light needs to travel
across the horizon, $\Delta >2R_S/c$. Now there is a natural time
period $\Delta_H$ associated with a black hole, namely that of its (imaginary)
Euclidean time axis (see the reviews [1,2] and
\cite{fr}): \begin{eqnarray} \Delta_H=\hbar\beta_H&=&8\pi\,M\,G/c^3=4\pi\,R_S/c
= 8\pi
 \frac{M}{m_{P}}\;t_{P}~, \\ & & t_{P}=\sqrt{\hbar G/c^5}
 =5.4\cdot 10^{-44}s~.\nonumber \end{eqnarray} It is tempting to identify
  $\Delta$ with
 $\Delta_H$. By doing so we obtain \begin{eqnarray}
 M_n&=&\frac{1}{2}\sqrt{n}\;m_{P}~,\\ A_n&=& 4\pi\,n\,l_{P}^2~.
  \end{eqnarray} Notice
 that the mass values (27) differ from those of the eq.(23) by the factor
 $\sqrt{\ln2/\pi}\approx 0.47$ only! \\  Let me give a -  not very
 satisfactory - plausibility argument why one may identify the period $\Delta$
 in real
 (i.e. Lorentzian) time $t$ with the period $\Delta_H$ in Euclidean time: A very
 simple  derivation \cite{ha} of  $\Delta_H$ starts from the Euclidean
 Schwarzschild line element \beq ds_E^2=
 (1-R_S/r)d(ict)^2+(1-R_S/r)^{-1}dr^2+r^2d\Omega^2~,\eeq which again has a
 coordinate singularity at $r=R_S$. But now $r\geq R_S$ always. Introducing
 the new coordinate \beq \rho=2R_S(1-R_S/r)^{1/2}\eeq the Euclidean
  line element (29) becomes \beq
  ds_{E}^2=\rho^2 d(\frac{ict}{2R_S})^2+(r/R_S)^4\,d\rho^2+r^2 d\Omega^2~.\eeq The
  first term has the form of polar coordinates in the plane with the angular
  variable $ict/2R_S$. If this has the period $2\pi$ then $it$ has the period
  $\Delta_H=4\pi\,R_S/c$. The argument no longer applies in the Lorentzian
  case. However, if t {\em is periodic from the start}, then it appears
   plausible to
  attribute to $t$ the period $\Delta_H$, too! It may be worth trying!
  \\ Suppose now that $n\gg 1$.
  Then we have $M_{n+1}-M_n = m_{P}^2/(8M_n)$ and can define the frequences
  $\omega_n$ by \beq \hbar
\omega_n=E_{n+1}-E_n=\frac{m_{P}^2c^2}{8M_n}=\frac{E_{P}}{4\sqrt{n}}~,~
E_{P}=1.2\cdot10^{22}\, \mbox{MeV}~.
   \eeq If we assume the frequency $\omega_n$ to be that of
   a gravitational or  an electromagnetic
   wave emitted in a transition of a system with mass $ M=M_{n+1}$ then this
    wave
   has the length \beq \lambda= \frac{2\pi\,c}{\omega_n}=16\pi\frac{M}{m_{P}}
   l_{P}\approx 80 \frac{M}{M_{\odot}}\mbox{ km },\eeq or the
   frequency \[\nu = c/\lambda \approx 3.8 \frac{M_{\odot}}{M}\mbox{ kHz}~~, \]
     where $M_{\odot}$ is
   the mass of the sun ($\approx 2\cdot 10^{30}\mbox{ kg}$). Thus, a black
    hole with a
   mass 10 times that of the sun would emit such waves with a  wave
   length of about $800$ km or a frequency of about 380 Hz!  \\
    If we define $\beta_n$ by $\beta_n=8\pi\,G\,M_n/(\hbar c^3)$ (see eq.
    (24))
    then we have the relation \beq \hbar \omega_n\,\beta_n=\pi~,\eeq which
   has a superficial similarity to Wien's displacement law $\hbar
   \omega_{max}\,\beta=2,82\ldots$, with $\omega_{max}$ as the frequency where
   Planck's distribution has its maximum for a given temperature
   $T=1/(k_B\beta)$. Whereas Wien's law has an essential physical content in
   connection with Planck's distribution for a given temperature $T$, the
   relation (34) appears mainly as a consequence of the definition of $\beta_n$!
   \section{Discussion} We have seen that, starting from the canonical
   quantisation of pure spherically symmetric gravity,  two assumptions,
    namely periodicity in
   time and identification of this period with the associated Euclidean time
   period $ \Delta_H$, lead to a mass spectrum which is very similar to those
   postulated or derived in other approaches. \\ Let me comment on the
    periodicity
   assumption first: As far as I can see it is not in contradiction to the
   basic
   elements which entered into the derivation of the Schroedinger eq.\ (11),
   because the second canonical variable $T$ ("time") does not appear at
    all in this
   equation. The properties of the time functional $T$ are, of course,
   affected and the details of its modification
    still have to be worked out. They are not essential for the main
    arguments above, because only the form of the reduced Hamiltonian (10)
    is important! \\ As the mass spectrum is no longer
   continuous, the scale transformations resulting from the
   commutation relations (9) cannot hold anymore and the scalar product (13) has
   to be changed. \\ In this connection there is
    another new feature: If we have free particles in a
   spatial box of length $L$ then this length is fixed for all wave functions,
   whereas the period $\Delta$ depends on the state characterized by $n$:
   \beq \Delta = \Delta_n
   =8\pi G\,M_n/c^3=8\pi\frac{M_n}{m_{P}}t_{P}=4\pi\sqrt{n}\,t_{P}\approx
   10^{-4}\frac{M}{M_{\odot}}s ~.\eeq (For $M_n\approx M_{\odot}$ one has
   $\sqrt{n} \approx 10^{38}$!) Thus, the wave functions (12) take the
   form \beq  c_n\,e^{\displaystyle -
    \frac{i}{\hbar}E_{P}\sqrt{n}\;\tau}~,~ 0\leq \tau \leq
   4\pi\,\sqrt{n}\; t_{P}~.\eeq Furthermore, postulating periodicity in time
   means that time translation invariance gets broken und formally energy is no
    longer
   conserved for the system. This, however, is intuitively obvious:
   Terminating a given state of an isolated system is only possible if some
   interactions with other systems are involved. We did not introduce such
   systems explicitly, but they are in the background of our time cutoff for
   the given system. \\ Next let me comment on the notion of entropy
   in this context: Contrary to the approach by Bekenstein and Mukhanov no
   properties of the entropy have to be used in order to derive the levels
  (27). Looking at the wave functions (36) the states they describe do not
  appear to be degenerate. However, this does not affect the formula (22),
   because
  the degeneracy $g(n)$ means something else, namely it counts the number of
  ways the level $n$ can be built up from $n=0$! So we still can introduce the
  entropy (22). However, we now have the strict inequality \beq S_n^{BM}/k_B <
  \frac{1}{4}\frac{A_n}{l_{P}^2}-\frac{\alpha}{4}~ \eeq for all $n$, because of the factor
  $\ln2/\pi\approx 0.22$ by which the relations (22) and (28) differ. This
  inequality is compatible with the general Bekenstein hypothesis \cite{be}
  \beq S/k_B \leq \frac{2\pi}{c\hbar}R\,E~,\eeq where $R$ is the radius
  of the smallest sphere that surrounds the system with energy $E$. \\
  If we relax the assumption $\Delta=\Delta_H$ and merely require \beq \Delta=
  \gamma R_S/c~, \eeq where $\gamma$ is of order 1 or larger, then we get
  instead of eq. (27) \beq M_n = \sqrt{\frac{\pi}{\gamma}}\sqrt{n}\;m_{P}~.
  \eeq For $\gamma =2$ the interval $\Delta$  coincides with
  Wald's \cite{wa2}
  "formation time" $\hbar/\kappa=\Delta_H/2\pi$, $\kappa$: surface gravity;
   and for $\gamma=3\sqrt{3}/2$ the interval $\Delta$ coincides with the
    characteristic
   luminosity attenuation time of a collapsing star \cite{at}. These examples
   show again that one should expect $\gamma$ to be not much larger than 1!
   Notice that the value of $\gamma$ affects the wave length (33) where the
   factor $16\pi$ after the second equality sign has to be replaced by
    $4\gamma$.
   \\ Finally I would like to compare the minimal area \beq a_0= A_1=4\pi\,
   l_{P}^2
   ~, \eeq resulting from eq. (28) with those of other authors: Bekenstein and
   Mukhanov have $a_0=4\ln2\,l_{P}^2$. Ashtekar and Lewandowski \cite{al}
   derive $a_0=\sqrt{3}\,l_{P}^2/4$.  \\ \\
   I thank S. Lau, F. Schramm and T. Strobl for discussions.
     \end{document}